# How to Track Online Service Level Agreement

Anuradha Rana[#1], Pratima Sharma[*2]

[#] Computer Science & Engineering, Maharshi Dayanand University
Advanced Institute of Technology & Management
Palwal, India

***Abstract*— SLA (Service level agreement) is defined by an organization to fulfil its client requirements, the time within which the deliverables should be turned over to the clients. Tracking of SLA can be done manually by checking the status, priority of any particular task. Manual SLA tracking takes time as one has to go over each and every task that needs to be completed.**

**For instance, you ordered a product from a website and you are not happy with the quality of the product and want to replace the same on urgent basis, You send mail to the customer support department, the query/complaint will be submitted in a queue and will be processed basis of its priority and urgency (The SLA for responding back to customers concern are listed in the policy). This online SLA tracking system will ensure that no queries/complaints are missed and are processed in an organized manner as per their priority and the date by when it should be handled. The portal will provide the status of the complaints for that particular day and the ones which have been pending since last week. The information can be refreshed as per the client need (within what time frame the complaint should be addressed).**

***Keywords*— SLAtracking,Priority,Request,Complaint, Scheduling,Resource, Quality**

## I. INTRODUCTION

Online Service Level agreement offers the benefit of cost-cutting to enterprises by online-allocation of work because cost depends on whenever and for how long the resources are required. Although cost-effective, this latest technology affects customary Security and trust mechanisms employed by these enterprises. Other than trust, there are few other sources that offer reimbursement in case there is a breach of mutual agreement. These may include court action or insurance protection. Of the different ways of establishing trust, the most important is security. Another component of online trust is reputation. Trust is also related to brand name and image.

## II. MEANING OF SERVICE LEVEL AGREEMENT

A **service-level agreement** (**SLA**) is a part of a service contract where a service is formally defined. In practice, the term *SLA* is sometimes used to refer to the contracted delivery time (of the service or performance).

A service-level agreement is an agreement between two or more parties, where one is the customer and the others are service providers. This can be a legally binding formal or an informal "contract" (for example, internal department relationships). Contracts between the service provider and other third parties are often (incorrectly) called SLAs – because the level of service has been set by the (principal) customer, there can be no "agreement" between third parties; these agreements are simply "contracts."

### A. *Motivating example*

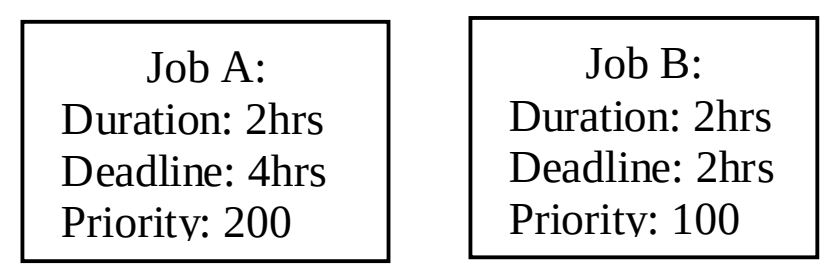

fig.1 An Example of this type of scheduling the problem the scheduler cannot manage. Out of the three job attributes of durtion, deadlines and priority the scheduler only uses the priority in making decisions.

In the example there are two jobs to be run. Both the jobs have same duration of 3 hrs. Job A has to be completed within 3hrs and Job B is worthwhile to the owner if and only if it is completed within 2hrs.Currently, job have no way to specify the temporal constraints. Instead the scheduler simply uses the priority value of jobs in determining which job is to run first. In this case, Job A would be run first because of its higher priority values, job B would miss it deadline. An SLA can be considered a legal binding contract that specifies the terms and levels of service. The parties of an SLA can be distinguished into providers and consumers of a service. The terms are agreed upon between service providers and consumers.

### B. *Performance Metrics used in SLA*

Service level agreements can contain numerous service performance metrics with corresponding service level objectives. A common case in IT service management is a call center or service desk. Metrics commonly agreed to in these cases include:

- **ABA** (Abandonment Rate): Percentage of calls abandoned while waiting to be answered.
- **ASA** (Average Speed to Answer): Average time (usually in seconds) it takes for a call to be answered by the service desk.

- **TSF** (Time Service Factor): Percentage of calls answered within a definite timeframe, e.g., 80% in 20 seconds.
- **FCR** (First-Call Resolution): Percentage of incoming calls that can be resolved without the use of a call back or without having the caller call back the helpdesk to finish resolving the case.
- **TAT** (Turn-Around Time): Time taken to complete a certain task.
- **MTTR** (mean Time to recover): Time taken to recover after an outage of service.

**Uptime** is also a common metric, often used for data services such as shared hosting, virtual private servers and dedicated servers. Common agreements include percentage of network uptime, power uptime, number of scheduled maintenance windows, etc. Many SLAs track to the Information Technology Infrastructure Library specifications when applied to IT services.

### *C. Why SLA is Important*

A good SLA is important because it sets boundaries and expectations for the following aspects of service provisioning.
• *Customer commitments*. Clearly defined promises reduce the chances of disappointing a customer. These promises also help staying focused on customer requirements and assuring that the internal processes follow the right direction.
• *Key performance indicators for the customer service*. By having these indicators established, it is easy to understand how they can be integrated in a quality improvement process. By doing so, improved customer satisfaction stays a clear objective.
• *Key performance indicators for the internal organizations*. A SLA drives internal processes by setting a clear, measurable standard of performance. Consequently, internal objectives become clearer and easier to measure.
• *The price of non-conformance*. If the SLA has penalties, non-performance can be costly. Therefore, by having penalties defined, the customer understands that the provider truly believes in its ability to achieve the set performance levels. It makes the relationship clear and positive.

## III. DESIGN OF SLA

### *A. Current State*

The process of SLA management in a customer care department is manual which is done by browsing the emails or queries received from the users/clients. There was no way to get the mails/queries prioritized. This leads to overlooking of urgent requests by the users when the inflow of requests is more thus having financial implications on the department for missing the SLA of that particular request.

### *B. Future State*

In order to prioritize the things, this project will help the department in getting the request which is nearing its expiration or SLA miss. The system is designed keeping in mind the priorities of different requests. The emails were sent to a common mailbox which was manually checked by a user accessing that mailbox. Below were the steps taken to get to the proposed solution:

1. The request or emails were categorized basis or urgency and type of request.

2. 4 priorities were defined for the requests which are Critical 1 hour, High 4 hour, Medium 1 day and Low 3 day.

3. Due Date was added for every request that is received and a counter was set in all the requests which will count the days since the day it was created.

4. Different queues were setup as per the priority of the request so they fall under their respective priority.

5. The number of requests which are about to miss SLA will show up on the web portal in the following manner.

### *C. Dummy Data used in Prototype*

In this implementation, we have used the experimental or dummy data as an example as this is related to **SLA Management System for MCF**

1. Complains like water supply, connection, map revision, general questions are lodged in MCF
2. We have divided these complaints in 5 categories which will be worked upon depending upon the time it takes for processing the same
3. This Online SLA system will track the status of complaints, assigned status of these complaints is as follows

   Critical, High, Medium, Low, Planned
4. Map revision, Tube well, Street Lights, etc will come under Planned Resolution as these require a lot of time in approval process and implementation. Similarly, different types are categorized as per the work
5. As user will logon and register any complaint or question, a unique number will be assigned to each complaint and as per the subject/ detail of the complaint the administrator will assign category to that.
6. After assigning the category, it will be kept in a queue for processing until it is assigned to executive who is responsible for handling that
7. Till the time it gets assigned to their id and name, the clock will monitor the Date the complaint was registered and the time. If it is Critical that needs to be done in 1 working day, high will be 2 working days, Medium will be 3 working days and Low will be 5 working days.

8. SLA management portal will monitor the time and show the complaint with its name and id which is about to miss the completion date
9. This portal will then help MCF to keep track of Critical, on-critical complains within completion time.
10. This will enhance people satisfaction and help MCF also to keep track of unassigned complaints.

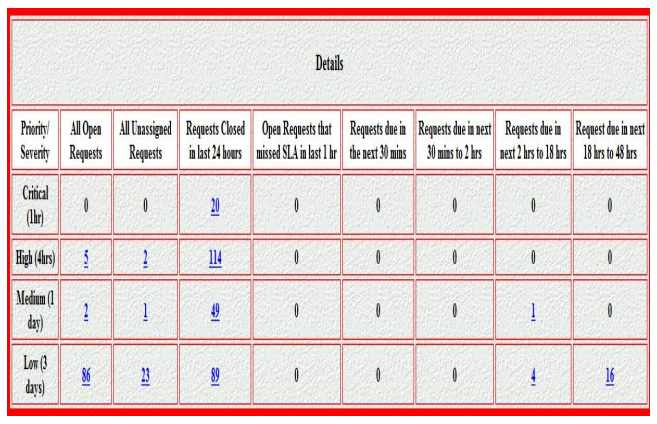

| Details | | | | | | | | |
|---|---|---|---|---|---|---|---|---|
| Priority/ Severity | All Open Requests | All Unassigned Requests | Requests Closed in last 24 hours | Open Requests that missed SLA in last 1 hr | Requests due in the next 30 mins | Requests due in next 30 mins to 2 hrs | Requests due in next 2 hrs to 18 hrs | Request due in next 18 hrs to 48 hrs |
| Critical (1hr) | 0 | 0 | 20 | 0 | 0 | 0 | 0 | 0 |
| High (4hrs) | 5 | 2 | 114 | 0 | 0 | 0 | 0 | 0 |
| Medium (1 day) | 2 | 1 | 49 | 0 | 0 | 0 | 1 | 0 |
| Low (3 days) | 86 | 23 | 89 | 0 | 0 | 0 | 4 | 16 |

Fig. 1 No of Request miss

## D. How Online SLA Works

Online SLA will allow the management to monitor the contribution of the various members in their team. To gauge exactly how well a team is performing overall, Online SLA allows you to capture and report specific data points for each individual within the team, thus providing a "snapshot" of performance.

## E. Benefits of using Online SLA include:

- Request is categorized on basis of priority
- Ability to identify and correct negative trends
- Measure efficiencies/inefficiencies
- Ability to generate detailed reports showing new trends
- Saves time compared to running multiple reports
- Gain total visibility of all systems instantly

## F. Flow Of Solution

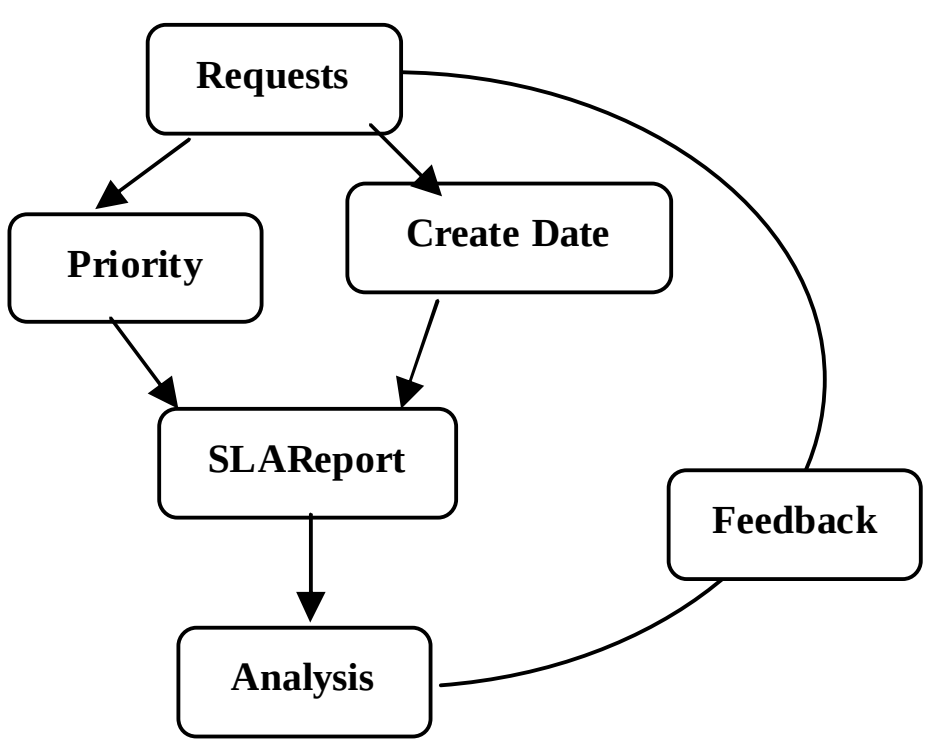

Fig 3. Flowchart of SLA reporting

## G. SLA Reporting

SLA Reporting is the main page. When you open Online SLA Tracking, on this page you will contain three buttons Change Priority Mapping, Prepare SLA File and Enter Request Data. One Blue Button that is for Show Report.

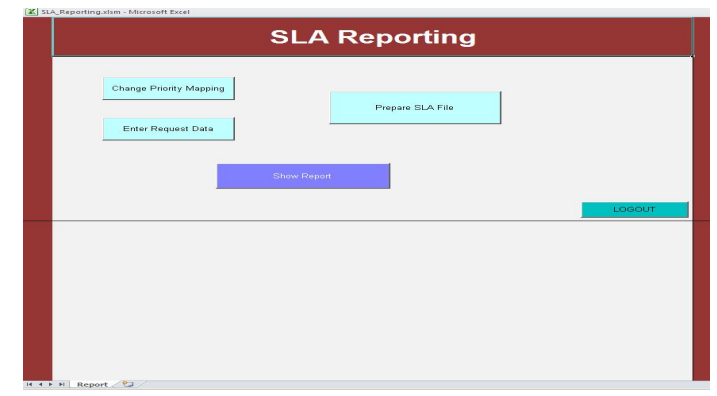

Fig.4. SLA Reporting

## H. Priority matrix

When we click on the button, **Change Priority Mapping**, it opens the Priority Mapping sheet,

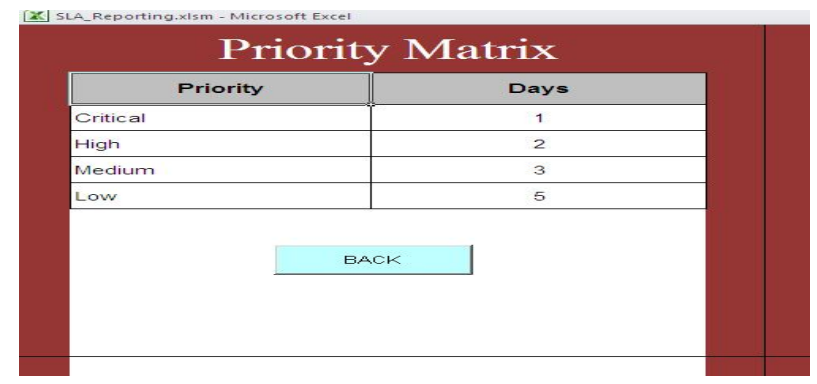

Priority Matrix

| Priority | Days |
|---|---|
| Critical | 1 |
| High | 2 |
| Medium | 3 |
| Low | 5 |

Fig 5. Priority Matrix

## I. Enter Request Data

When you click on Enter Request data, it displays the database.

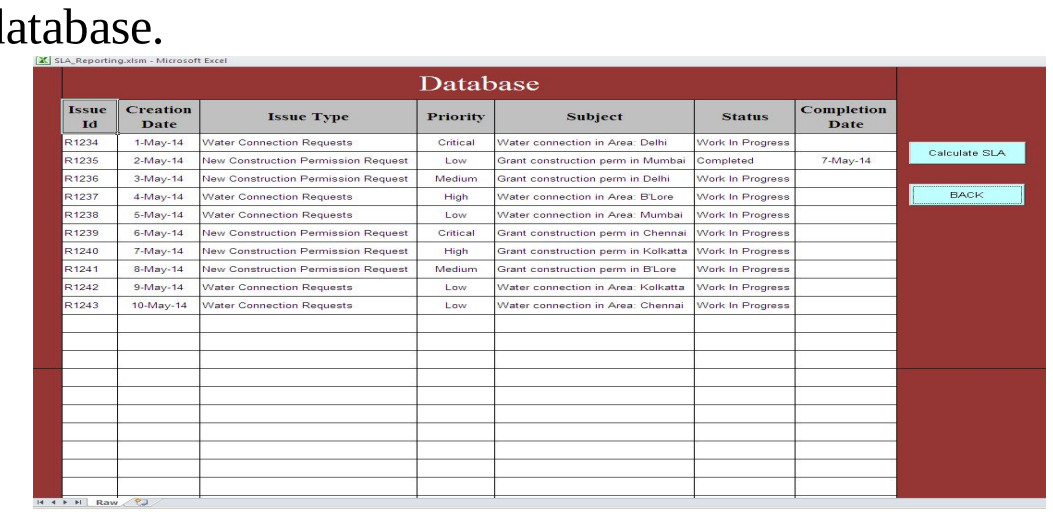

Database

| Issue Id | Creation Date | Issue Type | Priority | Subject | Status | Completion Date |
|---|---|---|---|---|---|---|
| R1234 | 1-May-14 | Water Connection Requests | Critical | Water connection in Area: Delhi | Work In Progress | |
| R1235 | 2-May-14 | New Construction Permission Request | Low | Grant construction perm in Mumbai | Completed | 7-May-14 |
| R1236 | 3-May-14 | New Construction Permission Request | Medium | Grant construction perm in Delhi | Work In Progress | |
| R1237 | 4-May-14 | Water Connection Requests | High | Water connection in Area: B'Lore | Work In Progress | |
| R1238 | 5-May-14 | Water Connection Requests | Low | Water connection in Area: Mumbai | Work In Progress | |
| R1239 | 6-May-14 | New Construction Permission Request | Critical | Grant construction perm in Chennai | Work In Progress | |
| R1240 | 7-May-14 | New Construction Permission Request | High | Grant construction perm in Kolkatta | Work In Progress | |
| R1241 | 8-May-14 | New Construction Permission Request | Medium | Grant construction perm in B'Lore | Work In Progress | |
| R1242 | 9-May-14 | Water Connection Requests | Low | Water connection in Area: Kolkatta | Work In Progress | |
| R1243 | 10-May-14 | Water Connection Requests | Low | Water connection in Area: Chennai | Work In Progress | |

Fig 6. Database(Enter Request Data)

## J. Calculate SLA

When we click on the button, **Calculate SLA**, it internally calculates the SLAs, based on the Priority & the Creation Date.

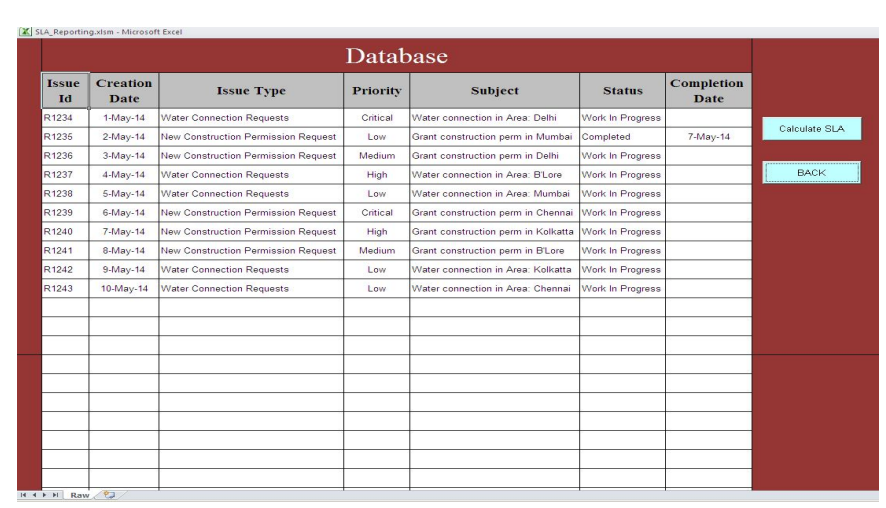

Database

| Issue Id | Creation Date | Issue Type | Priority | Subject | Status | Completion Date |
|---|---|---|---|---|---|---|
| R1234 | 1-May-14 | Water Connection Requests | Critical | Water connection in Area: Delhi | Work In Progress | |
| R1235 | 2-May-14 | New Construction Permission Request | Low | Grant construction perm in Mumbai | Completed | 7-May-14 |
| R1236 | 3-May-14 | New Construction Permission Request | Medium | Grant construction perm in Delhi | Work In Progress | |
| R1237 | 4-May-14 | Water Connection Requests | High | Water connection in Area: B'Lore | Work In Progress | |
| R1238 | 5-May-14 | Water Connection Requests | Low | Water connection in Area: Mumbai | Work In Progress | |
| R1239 | 6-May-14 | New Construction Permission Request | Critical | Grant construction perm in Chennai | Work In Progress | |
| R1240 | 7-May-14 | New Construction Permission Request | High | Grant construction perm in Kolkatta | Work In Progress | |
| R1241 | 8-May-14 | New Construction Permission Request | Medium | Grant construction perm in B'Lore | Work In Progress | |
| R1242 | 9-May-14 | Water Connection Requests | Low | Water connection in Area: Kolkatta | Work In Progress | |
| R1243 | 10-May-14 | Water Connection Requests | Low | Water connection in Area: Chennai | Work In Progress | |

**Fig 6.** Database (calculate SLA)

When we click on the button, **BACK**, it will again bring us back to the Main Page.

### K. *Create SLA File*

When we click on the button, **Prepare SLA File**, it will generate 2 types of files:

1. SLA file(s) – out_file.csv & out_file_detailed.csv
2. Settings file – Which provides the folder path of the SLA file(s). This Settings file will be used by the View exe later

### L. *SLA Files*

This Report view contains 2 parts:

1. Overview,

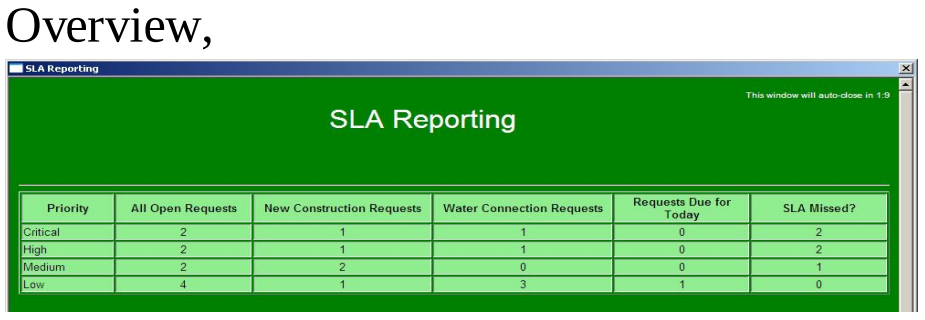

SLA Reporting

This window will auto-close in 1:9

| Priority | All Open Requests | New Construction Requests | Water Connection Requests | Requests Due for Today | SLA Missed? |
|---|---|---|---|---|---|
| Critical | 2 | 1 | 1 | 0 | 2 |
| High | 2 | 1 | 1 | 0 | 2 |
| Medium | 2 | 2 | 0 | 0 | 1 |
| Low | 4 | 1 | 3 | 1 | 0 |

Fig 7.Overview File

2. Detailed View

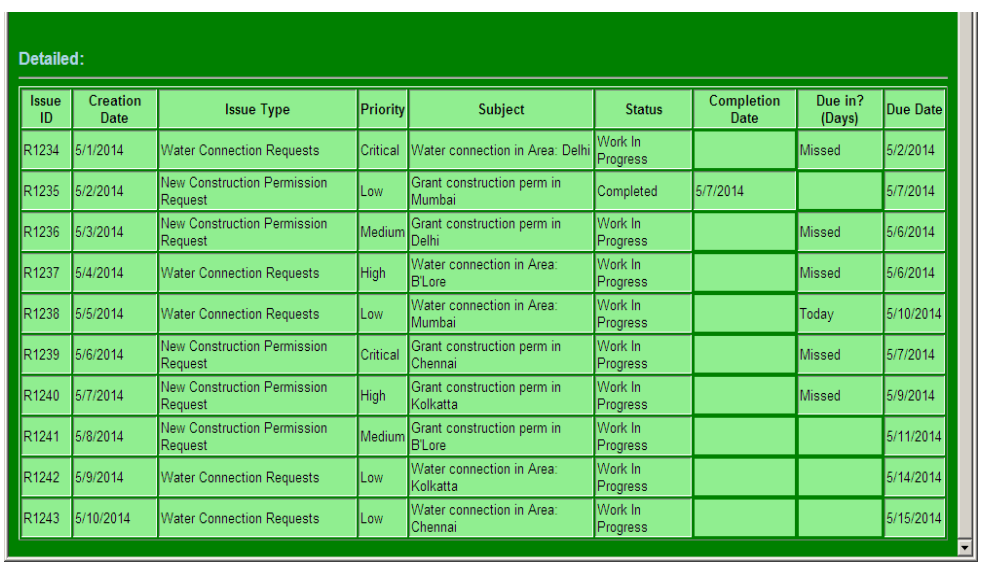

Detailed:

| Issue ID | Creation Date | Issue Type | Priority | Subject | Status | Completion Date | Due in? (Days) | Due Date |
|---|---|---|---|---|---|---|---|---|
| R1234 | 5/1/2014 | Water Connection Requests | Critical | Water connection in Area: Delhi | Work In Progress | | Missed | 5/2/2014 |
| R1235 | 5/2/2014 | New Construction Permission Request | Low | Grant construction perm in Mumbai | Completed | 5/7/2014 | | 5/7/2014 |
| R1236 | 5/3/2014 | New Construction Permission Request | Medium | Grant construction perm in Delhi | Work In Progress | | Missed | 5/6/2014 |
| R1237 | 5/4/2014 | Water Connection Requests | High | Water connection in Area: B'Lore | Work In Progress | | Missed | 5/6/2014 |
| R1238 | 5/5/2014 | Water Connection Requests | Low | Water connection in Area: Mumbai | Work In Progress | | Today | 5/10/2014 |
| R1239 | 5/6/2014 | New Construction Permission Request | Critical | Grant construction perm in Chennai | Work In Progress | | Missed | 5/7/2014 |
| R1240 | 5/7/2014 | New Construction Permission Request | High | Grant construction perm in Kolkatta | Work In Progress | | Missed | 5/9/2014 |
| R1241 | 5/8/2014 | New Construction Permission Request | Medium | Grant construction perm in B'Lore | Work In Progress | | | 5/11/2014 |
| R1242 | 5/9/2014 | Water Connection Requests | Low | Water connection in Area: Kolkatta | Work In Progress | | | 5/14/2014 |
| R1243 | 5/10/2014 | Water Connection Requests | Low | Water connection in Area: Chennai | Work In Progress | | | 5/15/2014 |

Fig 8. Detailed File

## IV. CONCLUSIONS

SLA monitoring will help MCF to build stronger relationships with the customers/people who come for their work , this will also increase the satisfaction not only for the MCF employees but for the people also. This project is a real time example of implementing the service level management as the problem which was foreseen month back can now be resolved with this solution. Every time people have to come at MCF they are hesitant for completing a task as they have to struggle for their complaints, requirements, other needs, etc but with this system/project if implemented and taken into consideration by MCF will help MCF and people to have better understanding of the work, processing of tasks on time, better relationship as this will eliminate third party indulgences for having work completed. If implemented on larger scale connecting all the facilities provided by MCF it will prove to be a boon for people.

### A. *Future Scope*

If this prototype is implemented connecting all the MCF of the different states, they will be able to communicate better and will have access to information of the people. The same system can be implemented in other government departments thus helping them to develop healthy relationships and better service to the people.

## ACKNOWLEDGMENT

I would like to place on record my deep sense of gratitude to **Dr. Deepti Sharma** Head of department of Computer Science and Engineering, for her generous guidance, help and useful suggestions.

I express my sincere gratitude to Asst. Prof. **Miss Pratima Sharma**, Dept. of Computer Science and Engineering, for her stimulating guidance, continuous encouragement and supervision throughout the course of present work. I also wish to extend my thanks to **Mr. Mahesh Singh** and other colleagues for attending my seminars and for their insightful comments and constructive suggestions to improve the quality of this research work.

I am extremely thankful to **Dr. R.S.Chaudhary** Director for providing me infrastructural facilities to work in, without which this work would not have been possible.

## REFERENCES

[1] A. Anderson, "An Introduction to the Web Services Policy Language", in Fifth IEEE Int'l Workshop on Policies for Distributed Systems and Networks, 2004.

[2] A. Andrieux, K. Czajkowski, A. Dan, K. Keahey, H.Ludwig, T. Nakata, J. Pruyne, J. Rofrano, S. Tuecke and M. Xu, "Web Services Agreement Specification (WS-Agreement)", Version 2006/09, World-Wide-Web Consortium (W3C), September 2006.
M. B. Chhetri, J. Lin, S. K. Goh, J. Yan, J. Y Zhang and R.Kowalczyk, A Coordinated Architecture for the Agent-based Service Level Agreement Negotiation ofWeb Service Composition, in Proc. of the AustralianSoftware Engineering Conference, Sydney, Australia, April 18 - 21, 2006. pp. 90-99.

[3] D. Greenwood, G. Vitaglione, L. Keller and M. Calist,Service Level Agreement Management with AdaptiveCoordination, in Proc. of International Conference onNetworking and Services (ICNS), July 16-19, 2006,Silicon Valley, USA. pp. 45.

[4] L. J. Jin, V. Machiraju and A. Sahai, "Analysis on ServiceLevel Agreement of Web Services", Research ReportHPL-2002-180, Hewlett-Packard Laboratories, June 2002.

[5] A. Keller and H. Ludwig, "The WSLA Framework:Specifying and Monitoring Service Level Agreements forWeb Services", Journal of Network and SystemsManagement, v.11 n.1, March 2003. pp. 57-81.

[6] H. Ludwig, A. Dan and R. Keamey, "Cremona: AnArchitecture and Library for Creation and Monitoring ofWS-Agreements", in Proc of 2nd ICSOC, New York,USA, November 15-18, 2004. pp. 65-74.

[7] H. Ludwig, A. Keller, A. Dan, R.P. King, and R. Franck,"Web Service Level Agreement (WSLA) LanguageSpecification", Version 1.0, International BusinessMachines Corporation (IBM), 2003.

[8] D. Ouelhadj, J. Garibaldi and J. MacLaren, "Amulti-agent infrastructure and a service level agreementnegotiation protocol for robust scheduling in gridcomputing". In Proc. of the European Grid Conference(EGC), February 14-16, 2005, Amsterdam, The Netherlands. pp. 651-660.

[9] A. Sahai, A. Durante and V. Machiraju, "TowardsAutomated SLA Management for Web Services",Research Report HPL-2001-310, Hewlett-Packard (HP)Laboratories, July 26, 2002.

[10] C. Sharp, J. Shewchuk, A. Vedamuthu, UmitYalyinalpand D. Orchard, "Web Services Policy Framework(WS-Policy)", Version 1.0, March 2006.

[11] "Five Key Points for Every SLA". Dell.com

[12] SLA - Global IP Network - NTT America - www.us.ntt.net

[13] Global Latency and Packet Delivery SLA - Verizon Business

[14] AT&T High Speed Internet Business Edition Service Level Agreement

[15] IT-Tude.com - Service Level Agreements